 \documentclass[preprint,prb,aps,draft,a4paper,superscriptaddress,nobibnotes,showpacs] {revtex4} 
\usepackage[final]{graphicx} 
\usepackage{dcolumn}         
\usepackage{bm}              
\usepackage{dcolumn}         
\usepackage{longtable} 
\usepackage{rotating}
\usepackage{tabularx}
\usepackage[usenames]{color}
\begin{document}
\title{ 
       The multi-state CASPT2 spin-orbit method
       }
\date{\today}
\author{Zoila Barandiar\'an}
\affiliation{Departamento de Qu\'{\i}mica, 
             Universidad Aut\'onoma de Madrid, 28049 Madrid, Spain}
\affiliation{Instituto Universitario de Ciencia de Materiales Nicol\'as Cabrera,
             Universidad Aut\'onoma de Madrid, 28049 Madrid, Spain}
             %
             %
\author{Luis Seijo}
\affiliation{Departamento de Qu\'{\i}mica, 
             Universidad Aut\'onoma de Madrid, 28049 Madrid, Spain}
\affiliation{Instituto Universitario de Ciencia de Materiales Nicol\'as Cabrera,
             Universidad Aut\'onoma de Madrid, 28049 Madrid, Spain}
 \begin{abstract}
We propose the multi-state complete-active-space second-order perturbation theory spin-orbit method (MS-CASPT2-SO) for electronic structure calculations.
It is a two-step spin-orbit coupling method that does not make use of energy shifts and
that intrinsically guarantees the correct characters of the small space wave functions that are used to calculate the spin-orbit couplings,
in contrast with previous two-step methods.
 \end{abstract}
\maketitle
\section{\label{SEC:intro}Introduction}
 In electronic structure two-step spin-orbit coupling methods,
 dynamic correlation is handled in the first step, 
 using the spin-free part of the Hamiltonian 
 and a large configurational space in variational or perturbational schemes.
 Then,
 spin-orbit coupling is handled in the second step,
 using an effective Hamiltonian
 and a small configurational space in spin-orbit configuration interaction (CI) calculations.
 In these methods,
 the effective Hamiltonian contains explicit energy shifts,
 which are a mean to transfer dynamic correlation effects from the first step
 to the second step in a simple and effective manner.~\cite{LLUSAR:96,VALLET:00,MALMQVIST:02}

 It has been found that the energy shifts of the spin-orbit free levels,
 which are driven by their energy order within each irreducible representation,
 can lead to anomalous results when 
 avoided crossings exist with significant change of character
 of the wave functions at each side,
 which take place at different nuclear positions in the large and in the small
 electronic configurational spaces.
 In these cases, the shifts must be assigned according to the characters of the wave 
 functions.~\cite{SANCHEZ-SANZ:10:b}
 This usually implies analyses of wave functions in both configurational spaces.
  
 The ultimate reason behind these problems,
 which are present in the available two-step methods,~\cite{LLUSAR:96,VALLET:00,MALMQVIST:02}
 is the different nature of the wave functions
 of the spin-free states in the large and in the small configurational spaces,
 so that, even when the avoided crossings do not exist or when they take place at
 the same nuclear positions in the large and in the small spaces,
 such a different nature makes the spin-orbit couplings calculated in the small space
 not as accurate (meaning as close to the spin-orbit couplings calculated in the large space) as desired.
 
Here, we propose the multi-state complete-active-space second-order perturbation theory spin-orbit method (MS-CASPT2-SO).
It is a two-step method that
does not make use of energy shifts and
that guarantees by construction the correct characters of the small space wave functions that are used to calculate the spin-orbit couplings.

\section{\label{SEC:method}The MS-CASPT2-SO method}

Let us assume we have  a many electron system with a Hamiltonian $\hat{H}$
which is made of the addition of a spin-free contribution, $\hat{H}^{SF}$,
and a spin-orbit coupling contribution, $\hat{H}^{SO}$:
\begin{eqnarray} \label{EQ:H}
  \hat{H}  = \hat{H}^{SF} + \hat{H}^{SO}
  \,.
\end{eqnarray}

In the spin-orbit free MS-CASPT2-SO method,
the procedure is initially the same as the MS-CASPT2 procedure:\cite{ZAITSEVSKII:95,FINLEY:98}
Several state-average complete-active-space self-consistent-field 
SA-CASSCF (or CASCI) states are calculated,
which define a reference configurational space called the $\cal P$ space.
Let us collect them in the row vector 
\mbox{$\underline{\Psi}^{CAS} = 
(\mid\Psi^{CAS}_1\rangle, \mid\Psi^{CAS}_2\rangle, \ldots, \mid\Psi^{CAS}_p\rangle)$},
where $p$ is the total number of SA-CASSCF states. 
These wave functions can be classified according to their values of 
spin quantum numbers and symmetry group irreducible representations and subspecies,
$SM_S\Gamma\gamma$,
but we will omit these labels here for simplicity.

In spin-orbit free MS-CASPT2 calculations,
the SA-CASSCF wave functions are used as a basis to calculate the matrix of a
spin-free second order effective Hamiltonian, $\hat{H}_{2nd}^{SF,eff}$, 
which is defined in Eq.~30 of Ref.~\onlinecite{FINLEY:98}
and depends only on the spin-free part of the Hamiltonian, $\hat{H}^{SF}$.
This matrix, 
which is $\underline{H}_{2nd}^{SF,eff, CAS} = 
\underline{\Psi}^{CAS \dagger}\,\hat{H}_{2nd}^{SF,eff}\,\underline{\Psi}^{CAS}$,
is diagonalized  in order to compute the 
MS-CASPT2 energies, as the eigenvalues,
and the modified SA-CASSCF (or CASCI) states, as the eigenfunctions:
\begin{eqnarray} \label{EQ:1}
  \underline{H}_{2nd}^{SF,eff, CAS}\,\underline{U}  = \underline{U}\,\underline{E}^{MS2}
  \,,
\end{eqnarray}
where $\underline{E}^{MS2}$ is a diagonal matrix with the MS-CASPT2 energies
${E}^{MS2}_1, {E}^{MS2}_2, \ldots, {E}^{MS2}_p$, as the diagonal elements
and $\underline{U}$ is a unitary transformation of the original SA-CASSCF wave functions
that preserves the $SM_S\Gamma\gamma$ values,
\begin{eqnarray} \label{EQ:2}
  \underline{\Psi}^{CAS^\prime} = \underline{\Psi}^{CAS}\,\underline{U}
  \,.
\end{eqnarray}
Obviously, the modified SA-CASSCF wave functions $ \underline{\Psi}^{CAS^\prime}$
also span the  $\cal P$ space.
What is important is that 
they are the appropriate zeroth-order basis for a second-order perturbation theory treatment of the dynamic correlation
that leads to the MS-CASPT2 energies~\cite{ZAITSEVSKII:95}
and 
they have the appropriate characters in correspondence with these energies. 

In the MS-CASPT2 spin-orbit calculations proposed here,
we can follow two alternative procedures that lead to the same result.
Both of them are based on the use of the spin-dependent effective Hamiltonian
that results from the addition of the spin-orbit coupling operator to the
spin-free effective Hamiltonian of the MS-CASPT2 method,
\begin{eqnarray} \label{EQ:Heff}
  \hat{H}^{eff} = \hat{H}_{2nd}^{SF,eff} + \hat{H}^{SO}  
  \,.
\end{eqnarray}

In the first procedure, 
which is a formal two-step procedure,
the regular spin-orbit free MS-CASPT2 calculation is completed and the 
{\it modified} SA-CASSCF wave functions $\underline{\Psi}^{CAS^\prime}$
are used as a basis for the matrix representation of 
$\hat{H}^{eff}$.
The resulting matrix,
\begin{eqnarray} \label{EQ:FIRST}
\underline{H}^{eff,CAS^\prime} = 
\underline{\Psi}^{CAS^\prime \dagger}\,\hat{H}^{eff}\,\underline{\Psi}^{CAS^\prime} =
\underline{E}^{MS2} + \underline{H}^{SO,CAS^\prime}
  \,,
\end{eqnarray}
with 
$\underline{H}^{SO,CAS^\prime} =
\underline{\Psi}^{CAS^\prime \dagger}\,\hat{H}^{SO}\,\underline{\Psi}^{CAS^\prime}$,
couples the modified SA-CASSCF states via spin-orbit coupling. 
(It couples different $SM_S\Gamma\gamma$ blocks
and it can be factorized according to double group irreducible representations.)
Its diagonalization gives the final energies and wave functions:
\begin{eqnarray} \label{EQ:3}
  \underline{H}^{eff,CAS^\prime}\,\underline{U}^{SO^\prime}  = 
  \underline{U}^{SO^\prime}\,\underline{E}^{MS2-SO}
  \,,
\end{eqnarray}
where $\underline{E}^{MS2-SO}$ is a diagonal matrix with the MS-CASPT2-SO target energies
${E}^{MS2-SO}_1, {E}^{MS2-SO}_2, \ldots, {E}^{MS2-SO}_p$ as the diagonal elements
and $\underline{U}^{SO^\prime}$ is a unitary transformation of the modified SA-CASSCF wave functions
that couples the $SM_S\Gamma\gamma$ values
and gives the target spin-orbit wave functions,
\begin{eqnarray} \label{EQ:5}
  \underline{\Psi}^{MS2-SO} = \underline{\Psi}^{CAS^\prime}\,\underline{U}^{SO^\prime}
  \,.
\end{eqnarray}

Alternatively, in the second procedure, which is a formal one-step procedure,
the {\it original} SA-CASSCF wave functions $\underline{\Psi}^{CAS}$
are used as the basis for the matrix representation of 
$\hat{H}^{eff}$.
In order to do this,
the regular spin-orbit free MS-CASPT2 calculation does not  need to be completed,
but only the computation of the  $\underline{H}_{2nd}^{SF,eff, CAS}$  matrix used in Eq.~\ref{EQ:1}, 
plus the addition of the matrix of $\hat{H}^{SO}$ in this basis
($\underline{H}^{SO, CAS} = 
\underline{\Psi}^{CAS \dagger}\,\hat{H}^{SO}\,\underline{\Psi}^{CAS}$):
\begin{eqnarray} \label{EQ:SECOND}
  \underline{H}^{eff, CAS} = 
  \underline{\Psi}^{CAS \dagger}\,\hat{H}^{eff}\,\underline{\Psi}^{CAS} =
  \underline{H}_{2nd}^{SF,eff,CAS} + \underline{H}^{SO, CAS}
  \,.
\end{eqnarray}
Its diagonalization gives the same target energies and wave functions as the first procedure,
\begin{eqnarray} 
  \underline{H}^{eff,CAS}\,\underline{U}^{SO}  = 
  \underline{U}^{SO}\,\underline{E}^{MS2-SO}
  \,,
\end{eqnarray}
with
\begin{eqnarray} 
  \underline{\Psi}^{MS2-SO} = \underline{\Psi}^{CAS}\,\underline{U}^{SO}
  \,
\end{eqnarray}
and $\underline{U}^{SO} = \underline{U}\,\underline{U}^{SO^\prime}$.

The present method is closely related with the 
restricted active space state interaction approach with spin-orbit coupling
of Ref.~\onlinecite{MALMQVIST:02}, SO-RASSI.
The latter, when it is used in a SA-CASSCF/MS-CASPT2 context,
corresponds to diagonalizing the mixed effective Hamiltonian matrix
$\underline{E}^{MS2} + \underline{H}^{SO,CAS}$,
with the spin-orbit free part of Eq.~\ref{EQ:FIRST}
and the spin-orbit  coupling part of Eq.~\ref{EQ:SECOND}. 
The results of both approaches are expected to be similar when
the CAS and the CAS$^\prime$ wave functions (Eq.~\ref{EQ:2}) are also similar,
which is a common case. 
Differences should show up when the two sets of wave functions are not so similar in
a one-to-one basis, for instance when the dynamic correlation switches the order of states.
This is the basic advantage of the present method.
However, we must note that the SO-RASSI method can be used together with general
single-state and state-average RASSCF and CASSCF 
plus RASPT2 and single-state and multi-state CASPT2 spin-orbit free frameworks,
whereas the present one can only be used in the spin-orbit free framework
of SA-CASSCF plus MS-CASPT2 calculations. 

Let us now justify why $\hat{H}^{eff}$ (Eq.~\ref{EQ:Heff}) 
is the effective Hamiltonian of choice in the MS-CASPT2-SO method.
For this purpose,
we recall that the basic idea of two-step methods is 
to use a spin-orbit effective Hamiltonian  
made of a spin-free effective Hamiltonian
(usually $\hat{H}^{SF} + \hat{H}^{shift}$) 
plus the spin-orbit coupling operator $\hat{H}^{SO}$,
and to choose
the spin-free effective Hamiltonian
 by imposing the requirement that,
when used in the small space $\cal P$ of the second step,
it has the same eigenvalues that $\hat{H}^{SF}$ has 
in the large space $\cal G$ of the first step.~\cite{LLUSAR:96} 
In the particular case in which
the first step is a MS-CASPT2 calculation
and the small space of the second step is defined by the SA-CASSCF wave functions,
$\hat{H}_{2nd}^{SF,eff}$  is a spin-free effective Hamiltonian
that fulfills such a condition.
In consequence, $\hat{H}^{eff}$ (Eq.~\ref{EQ:Heff})
is the proper spin-orbit effective Hamiltonian.


\section{\label{SEC:conclusions}Conclusion}
A two-step spin-orbit coupling method for 
multi-state complete-active-space second-order perturbation theory calculations 
MS-CASPT2 is proposed
which does not make use of energy shifts.
It intrinsically guarantees the correct characters of the small space wave functions used to calculate the spin-orbit couplings,
in contrast with previous two-step spin-orbit coupling methods,
where it has to be checked externally.

 \acknowledgments
This work was partly supported by
a grant from Ministerio de Ciencia e Innovaci\'on, Spain
(Direcci\'on General de Programas y Transferencia de Conocimiento MAT2008-05379/MAT).

%

\end{document}